\documentclass[aps,prl,twocolumn,color,psfig,epsf,superscriptaddress]{revtex4-2}
\usepackage[english]{babel}
\usepackage[T1]{fontenc}
\usepackage{tikz}
\usepackage{amssymb,amsmath,amsfonts}
\usepackage{textcomp}
\usepackage{braket}
\usepackage{nicematrix} 
\usepackage{graphicx,color}
\usepackage[dvipsnames]{xcolor}
\usepackage[utf8x]{inputenc}
\usepackage{bm}
\usepackage{mathbbol}
\usepackage{float}
\usepackage{comment}
\usepackage[linkcolor = blue, citecolor = red, urlcolor = blue, colorlinks = true]{hyperref}
\usepackage{tabularx}
\newcommand{\1}{{1\hspace*{-0.5ex} \textrm{l} \hspace*{0.5ex}}}
\newcommand{\beq}{\begin{equation}}
\newcommand{\eeq}{\end{equation}}
\newcommand{\bea}{\begin{eqnarray}}
\newcommand{\eea}{\end{eqnarray}}

\begin{document}

\title{Braiding and exchange statistics of liquid crystalline Majorana quasiparticles}
\author{ A.~I. T\'oth$^{1,*}$, G. Negro$^{1,*}$, A.~D. Huxley$^{1}$, D. Marenduzzo}
\affiliation{SUPA, School of Physics and Astronomy, University of Edinburgh, Edinburgh EH9 3FD, UK\\
$^{*}$Equal contributions}

\begin{abstract}
Liquid crystalline defects in 3D can be viewed as geometric spinors, whose emergent properties are reminiscent of those of topological excitations in quantum condensed matter, such as Majorana quasiparticles. However, it is unclear how deep this analogy is, and whether this is a purely mathematical mapping, or it extends to key physical features, such as the exchange statistics or braiding behaviour. To address this question, here we consider a simple pattern made up of four nematic Majorana-like defect profiles, and ask how the defect profiles change as we braid them repeatedly around each other. Surprisingly, we find that in a large range of parameter space the defect profiles behave as classical analogues of non-Abelian 
anyons, which can be described in our case by defect bivectors moving on a Bloch-like hemisphere. Elastic interactions and dynamical effects enhance the complexity of the gates which can be performed by braiding these quasiparticles, making these liquid crystalline spinors promising candidates as components of topological computers. 
\end{abstract}

\maketitle

Solitons in condensed matter are low-energy excitations with which exhibit 
particle-like properties. Examples include domain walls, classical or quantum vortices~\cite{ivanov2001,oneill2022,hall2026,kleckner2013}, skyrmions~\cite{fert2017}, 
hopfions and torons~\cite{wu2022,Ackerman2013}. Liquid crystals provide a classical platform to create topological defects which behave as such particles~\cite{mermin1979,head2024,johnson2025}. Comet defects in active nematics 
can be viewed as self-propelled particles~\cite{keber2014}, whereas nematic or cholesteric 
disclinations exhibit features of Majorana and Weyl fermions~\cite{head2024,johnson2025}, providing a realisation in soft matter that is complementary to those of quantum condensed matter~\cite{alicea2012,bernevig2013}.

In quantum condensed matter physics, the braiding of quasiparticles is an important topic, as it reveals the exchange statistics as fermions, bosons or anyons, and simultaneously paves the way for applications in quantum computation~\cite{nayak2008,wang2010}. Liquid crystalline defect patterns 
can be viewed as geometric spinors, as they can be represented as a flag-pole, where the pole is the axis along which the director field performs a $\pi$ rotation (the defect bivector~\cite{johnson2025}), while the flag denotes the in-plane orientation of the defect profile 
which can be interpreted as a global geometric phase~\cite{kos2022}. In 3D nematics, two-dimensional profiles of opposite winding number, which are antidefects of each other, can be transformed into one another by a smooth transformation, hence they can be viewed as Majorana-like quasiparticles at rest~\cite{head2024} -- in our context, they are best viewed as analogues of Majorana zero modes~\cite{ivanov2001}. With respect to superconductors and superfluids, which can also contain topological excitations and Majorana modes~\cite{ivanov2001,mermin1979,salomaa1987}, the spatiotemporal scales of liquid crystal patterns are usually much larger ($\mu$m and ms), which can simplify their experimental study and manipulation~\cite{wu2022,head2024,johnson2025}. Additionally, topological quasiparticles can be realised in liquid crystals with readily available materials such as 5CB and E7 mixtures~\cite{wu2022}.
 
Here we ask what rules underpin the braiding of 
nematic defects. We do so by studying a quasi-two-dimensional system where defect profiles akin to Majorana zero modes are created on a plane by local melting of the liquid crystal, for instance modelling local heating by a laser. The system is quasi-2D because, while the defects move on the plane, the underlying director field can tilt out of the plane: this plays a key role. 
By simulating the liquid crystal hydrodynamic equations of motion with a hybrid lattice Boltzmann algorithm~\cite{marenduzzo2007,carenza2019}, we monitor the evolution of the director field pattern that arises when the defect quasiparticles are braided. 
In $p-$atic liquid crystals, defect braiding has been shown to produce path-dependent transformations of the underlying field configuration~\cite{mietke2022}, while in metamaterials non-Abelian braiding arises in mode space~\cite{chen2025}. Here we find that nematic defect braiding instead realises an emergent qubit-like system on the Bloch hemisphere, which is closely related geometrically to Ising anyons associated with Majorana zero modes~\cite{bombin2010}.
Some details differ, as in our case braiding can be affected by nematic elasticity and kinetic effects; additionally, the classical framework we use does not yield 
unitary dynamics~\cite{kos2022}. Our results suggest 
that braiding nematic defects 
may provide a practical pathway to classical topological computing. 

\begin{figure}
    \centering
    \includegraphics[width=1.0\columnwidth]{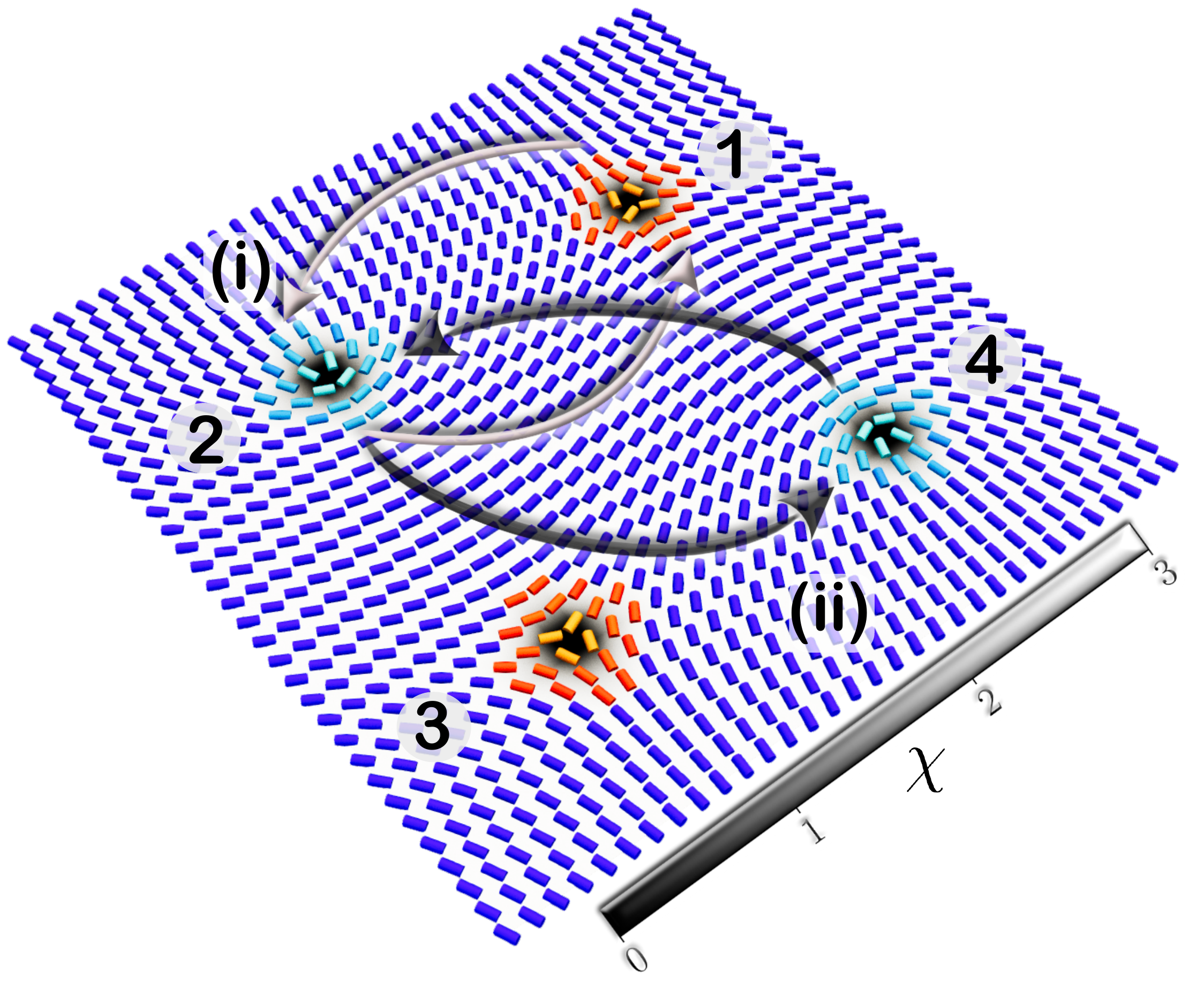}
    \caption{Sketch of the $4$-defect Majorana square, with alternating $-1/2$ (orange) and $+1/2$ (cyan) defects. The overall pattern 
    behaves as a ``nematic bit'', which can be described by a geometric spinor on the Bloch sphere (see text; in this starting configuration the spinor is aligned along $\pm \hat{\bf z}$). We also sketch the trajectories simulated when exchanging defects $1$ with $2$ (i), with opposite winding number (i), and $2$ with $4$ (ii), with equal winding number. Black spots indicate regions with $\chi = 0$, where the simulated laser is applied.}
    \label{fig1}
\end{figure}

To model the dynamics of liquid crystalline quasiparticles, we study the 
time evolution of the nematic ${\bm Q}$ tensor. 
The equilibrium properties of the system are described by a free energy, whose density, in dimensionless form, $\tilde{f}=f/A_0$ (with $A_0$ a typical energy scale quantifying the cost of local melting into the isotropic phase) is 
\begin{eqnarray}\label{eqn:3}  
    \tilde{f} = & &  \dfrac{1}{2} \left( 1-\frac{\chi}{3} \right) Q_{\alpha\beta}^2  -\frac{\chi}{3} Q_{\alpha\beta}Q_{\beta\gamma}Q_{\gamma\alpha}  \\ \nonumber
    &+& \frac{\chi}{4} (Q_{\alpha\beta}^2)^2  
    +\frac{l_n}{2}(\nabla \bm{Q})^2 .  
\end{eqnarray}
In Eq.~\eqref{eqn:3}, 
$l_n$ is the nematic correlation length~\cite{degennes1993}, which equals to $\sqrt{K/A_0}$ with $K$ the elastic constant, while $\chi$ is a temperature-like parameter that drives the isotropic-nematic transition, occurring for $\chi>\chi_{cr}=2.7$~\cite{degennes1993}. 
To simulate the effect of laser heating on the liquid crystals, we consider a spatially dependent value of $\chi$, $\chi({\mathbf x})$, which is equal to $\chi_0=0$ inside laser wells, modelled as circular regions of radius $R$, and equal to $\chi_1=3$ outside these, stabilising the defect pattern shown in Fig.~\ref{fig1}. In what follows, we vary the correlation length $l_n$ in our simulations (by varying $K$ at fixed $A_0=1$).

The dynamic of the orientational order is governed by the following equation
\begin{equation}\label{eqn:Q}
D_t {\bm Q}=-\Gamma\left[\frac{\delta {\mathcal F}}{\delta {\bm Q}}+({\bm I}/3){\rm Tr}\frac{\delta {\mathcal F}}{\delta {\bm Q}}\right],  
\end{equation}
where the term in square brackets is the molecular field, which drives the system to a local equilibrium, while $\Gamma$ is a relaxation constant, which is inversely proportional to the rotational viscosity $\gamma_1$~\cite{edwards1991}. The latter determines the relaxation timescale $\tau\sim l^2/{(\Gamma l_n^2 A_0)}$, where $l$ is a typical lengthscale -- in our case typically the defect distance or system size.
The differential operator $D_t$ equals $\partial_t$ in the absence of flow, which we consider in the main text. Inclusion of flow leads to qualitatively similar results, as discussed in~\cite{SI}. We integrated  Eq.~\eqref{eqn:Q} (and the Navier-Stokes equation when included~\cite{SI}) using a 
hybrid lattice Boltzmann approach~\cite{marenduzzo2007,carenza2019}. 
More details on the methods and a full parameter list are given in~\cite{SI}. 

\begin{figure*}
    \centering
    \includegraphics[width=2.0\columnwidth]{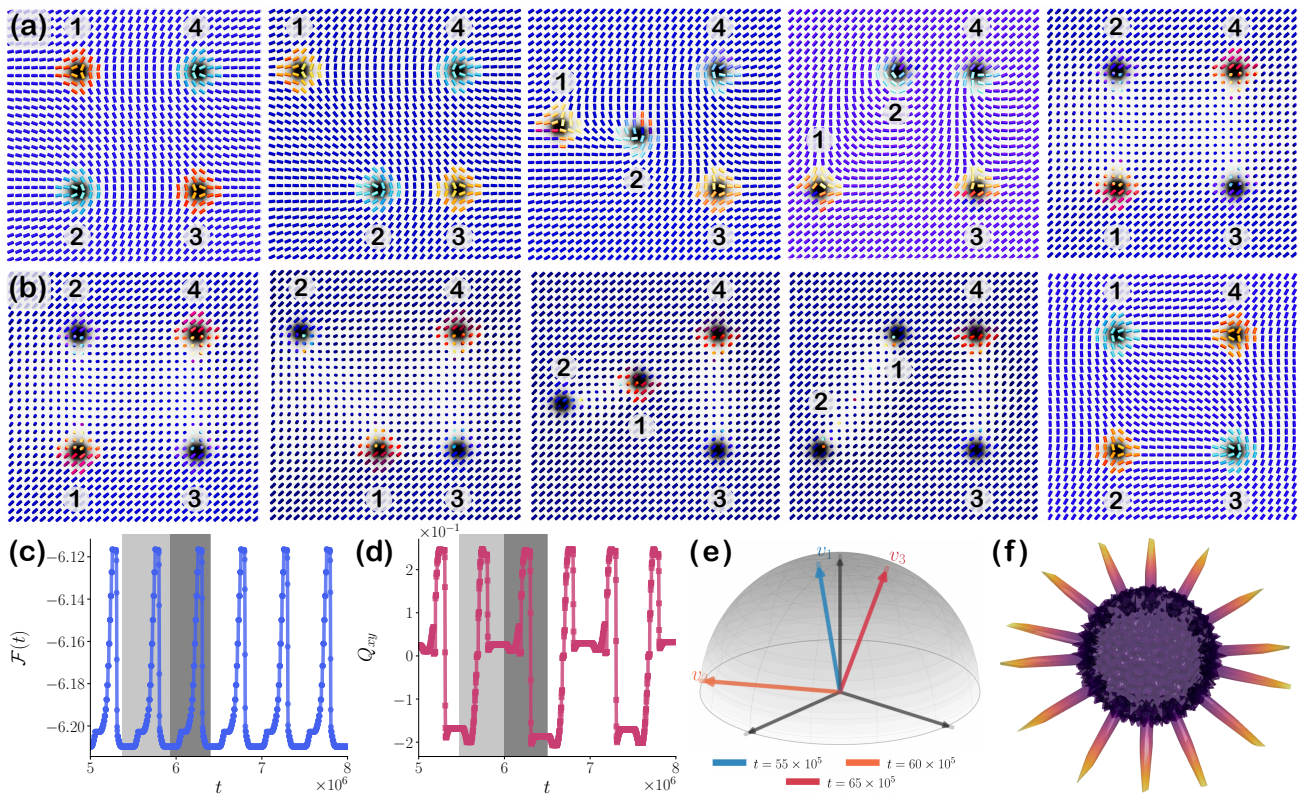}
    \caption{(a,b) Liquid crystal patterns associated with one cycle of the repeated exchange between defects $1$ and $2$ in the Majorana square. Snapshots in the first row (a) correspond to the first exchange, and those in the second row (b) to the second exchange, which returns the system to a configuration which is related to the starting one by a $90^{\circ}$ in-plane overall rotation.  
    Defects of winding 
    $-1/2$ (orange) and $+1/2$ (cyan) have been tracked using the local charge density in Eq.~(\ref{defDij}), hence the small variation in the color of the background nematic pattern when the director rotates out of the plane. 
    (c) Free energy versus time, showing that the states obtained after each exchange have the same free energy.  (d) Value of $Q_{xy}$ in the middle of the square as a function of time. 
    Light- and dark-gray shading marks the time windows of the first (a) and second (b) exchange events, respectively.
    (e) Plot of the defect bivector $\Omega$ (corresponding to the axis of the  Majorana square bivector) for configurations corresponding to the three successive square configurations in (a,b). 
    (f) Superposition of spherical harmonics (up to $99$th order) in a spherical harmonics expansion of the orbits on the Bloch sphere, corresponding to $64$ exchanges of the first two defects.
    }     
    \label{fig2}
\end{figure*}

\begin{figure}
    \centering
    \includegraphics[width=\linewidth]{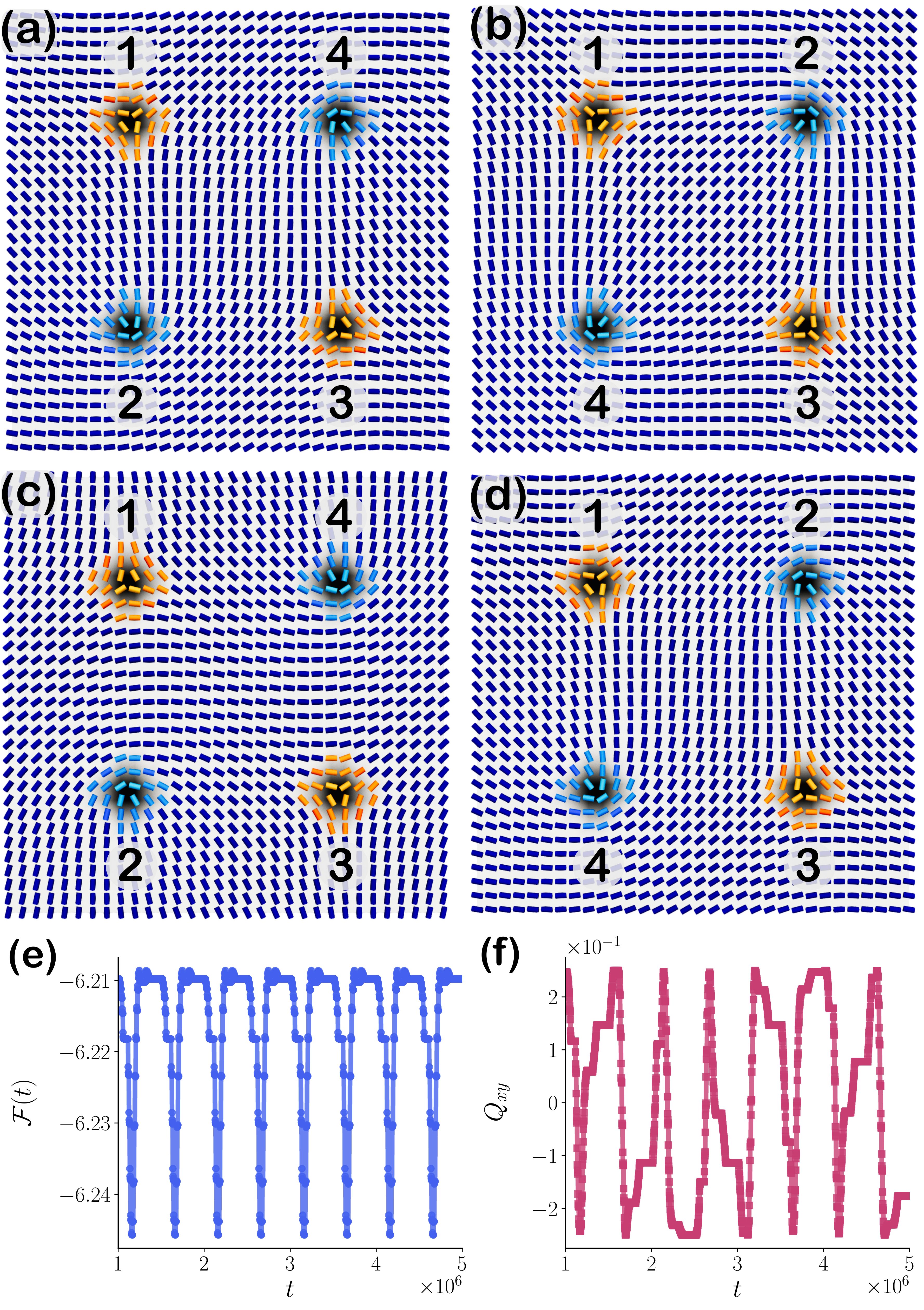}
    \caption{(a-d) Liquid crystal patterns following repeated exchange between defects $2$ and $4$ in the Majorana square. 
    (e) Free energy versus time, showing that after each exchange the states have the same, or nearly the same, free energy.  (f) Value of $Q_{xy}$ in the middle of the square as a function of time.  
    }     
    \label{fig3}
\end{figure}

We study a ``Majorana square'', with $4$ defect profiles initially at the four vertices of a square of size $d$, with alternating winding number 
[Fig.~\ref{fig1}]. The Majorana square lies inside a larger square domain of size $\mathcal{L}$ with periodic boundary conditions.  
The director field [Fig.~\ref{fig1}], ${\mathbf n}$, determines the initial condition used in simulations for the ${\bf Q}$ tensor, as we initially set $Q_{\alpha\beta}=q\left(n_{\alpha}n_\beta-\tfrac{\delta_{\alpha\beta}}{3}\right)$ (uniaxial approximation), with $q$ the magnitude of order, which 
equals $1/2$ for $\chi=3$~\cite{edwards1991} (and $0$ for $\chi=0$). 
Fig.~\ref{fig1} also shows the well labelling, as well as the exchange or braiding trajectories~\cite{SI} which we use for the simulations discussed in Figs.~\ref{fig2},\ref{fig3}.

We first analyse the dynamics accompanying the repeated exchange (or braiding) of defects $1$ and $2$, following the counterclockwise pathway sketched in Fig.~\ref{fig1}(i). Fig.~\ref{fig2} and Suppl. Movie 1 show the behaviour for $l_n=0.2$ and $d=16$ (for other parameter values, see~\cite{SI}). As the two defects exchange, the whole liquid crystal rearranges, with the director field twisting out of the plane into the third ($z$) dimension [Fig.~\ref{fig2}(a)]. A second exchange returns the initial condition, but up to an overall rotation [Fig.~\ref{fig2}(b)]; therefore in the sequence shown in Figs.~\ref{fig2}(a,b) we need four exchanges of the particles to get back exactly to the initial configuration [Suppl. Movie 1 and Fig.~\ref{fig2}(d)], a behaviour which likens our classical Majorana zero modes to anyonic quasiparticles~\footnote{In our simulations, the configuration typically returns to the same position after either two or four exchanges, depending on elasticity and noise. 
In either case, the behaviour upon exchange depends on which defects are braided (Fig.~\ref{fig3}), as for non-Abelian anyons.}. 

The free energy of the system, computed by summing the densities in Eqs.~(\ref{eqn:3}) over the simulation domain, oscillates during the braiding [Fig.~\ref{fig2}(c)], and is equal (within numerical accuracy) following each exchange, showing that the two configurations before and after an exchange are degenerate, or nearly so. The data also show that there is a barrier separating the two states, so that active braiding is required to reach one state from the other. 

To understand the behaviour of the Majorana square upon particle exchange, we define a defect bivector $\hat{\mathbf{\Omega}}$ as in~\cite{johnson2025}. This is built by starting from the disclination tensor \cite{head2024,schimming2022}, 
\begin{align}\label{defDij}
    D_{ij}=\epsilon_{i\mu\nu}\epsilon_{jlk}\partial_l Q_{\mu\alpha}\partial_k Q_{\nu\alpha}
\end{align}
where $i,j,k,\alpha,\mu,\nu$ are tensor indices and where the Einstein summation convention of repeated indices has been used. [Note that $D_{zz}$ is also the in-plane topological charge density~\cite{schimming2022}.] The defect bivector 
$\hat{\mathbf{\Omega}}$ is found by using singular value decomposition near the corresponding defect, to write $D_{ij} = s(\mathbf{r})\hat{\Omega}_{i}\hat{T}_j$, 
where $s(\mathbf{r})$ is a positive scalar field that is maximum at the disclination core, and $\hat{\mathbf{T}}$ is the local tangent to the disclination line. We can also write
$\hat{\mathbf{\Omega}}=\hat{\Omega}_x e_{32}+\hat{\Omega}_y e_{13}+\hat{\Omega}_z e_{21}$, with $e_{1,2,3}$ the generators of the Clifford algebra $Cl(3,0)$~\cite{johnson2025,hestenes2003,francis2005}, and $e_{21}=e_2e_1$, $e_{13}=e_1e_3$, $e_{32}=e_3e_2$ the three bivectors squaring to $-{\1}$, which can be thought of as the three quaternions $i$, $j$, $k$. This representation is useful to draw analogies to quantum condensed matter as it elucidates the spinor nature of the defect  profile~\cite{head2024,johnson2025,kos2022,copar2014}. 

While each of the four Majorana zero modes in the square has an associated bivector ${\hat{\mathbf{\Omega_i}}}$, with $i=1,\ldots, 4$, in practice 
all bivectors lock along the same direction due to elastic interactions between them mediated by the liquid crystalline medium. 
Additionally, the four defect bivectors have effective antipolar order, so that they sum to $0$, such that the topological charge of the system vanishes. As a result, the whole square is described by a single bivector direction, 
$\pm \hat{\mathbf{\Omega}}=\pm\hat{\mathbf{\Omega}}_1$. This square direction can be viewed as a bidirectional vector on the Bloch sphere,
or a unit vector on the Bloch hemisphere, and is equivalent to a single ``nematic bit''~\cite{kos2022}, or a single qubit. In Fig.~\ref{fig2}(e), we show how the defect bivector moves on the Bloch sphere after each exchange: the average rotation is $\theta\simeq 76^{\circ}$. 
The entire orbit of the square bivector consists of a discrete set of points along a large circle on the Bloch sphere [Fig.~\ref{fig2}(f)]. 


We next consider what happens when we braid defects $2$ and $4$, following the exchange trajectory sketched in Fig.~\ref{fig1}(ii). In this case we exchange two defects profiles with the same winding number. 
Fig.~\ref{fig3}(a) and Suppl. Movie 5 show the liquid crystal patterns formed dynamically during the exchange. Interestingly, in this case the director field remains on the plane, so that the defect bivectors is constant along $\pm \hat{z}$ at all times, and does not move along the Bloch sphere. The profiles rotate though, which is equivalent to a change in the global phase of the nematic bit.  
Fig.~\ref{fig3}(a-d) shows that braiding rotates the director profile inside the Majorana square:  the system returns to (approximately) the starting configuration after six exchanges, although in general the rotation angle depends on $l_n/d$. 
As in Fig.~\ref{fig2}, the free energy does not change once the defects exchange position 
[Fig.~\ref{fig3}(e)], as our system is symmetric with respect to global rotations. 

It is useful to compare and contrast 
the behaviour of our Majorana square with the braiding of Majorana zero modes~\cite{simon2023,SI}. 
Majorana zero modes can be described by non-Abelian ``Ising anyons'', $\sigma$, which can further combine pairwise to give either the vacuum, $\1$, or a spinless 
fermion, $\psi$. The composition between anyons and quasiparticles is determined by the following fusion algebra 
\begin{equation}
    \sigma \times \sigma = \1 + \psi \,; \, \sigma\times \psi = \sigma \, ; \, \psi\times \psi =\1 \, 
\end{equation}
whereas fusing $\1$ with any member of the algebra leaves that member unchanged. Four $\sigma$ 
anyons can set up an ``even parity'' system, which has trivial global topological charge. The even parity sector consists of linear superpositions of the vacuum state $\ket{0}$, which has no fermions, 
and the state $\ket{1}$, which has two $\psi$ fermions.  
The non-trivial $\sigma$ braiding matrix can be represented 
by~\cite{SI}
\begin{equation}
    B = \frac{1}{\sqrt{2}} 
    \begin{pmatrix}
        1 & -i \\
        -i & 1
    \end{pmatrix} = e^{-i \frac{\pi}{4}\sigma_x} 
\end{equation}
which corresponds to a Bloch rotation of $90^{\circ}$ 
vector rotation, or a $45^{\circ}$ 
spinor rotation, around the $x$ axis. This is a Clifford gate in quantum computing, or a member of the Clifford octahedral rotation subgroup.  

\begin{figure}
    \centering
    \includegraphics[width=0.9\columnwidth]{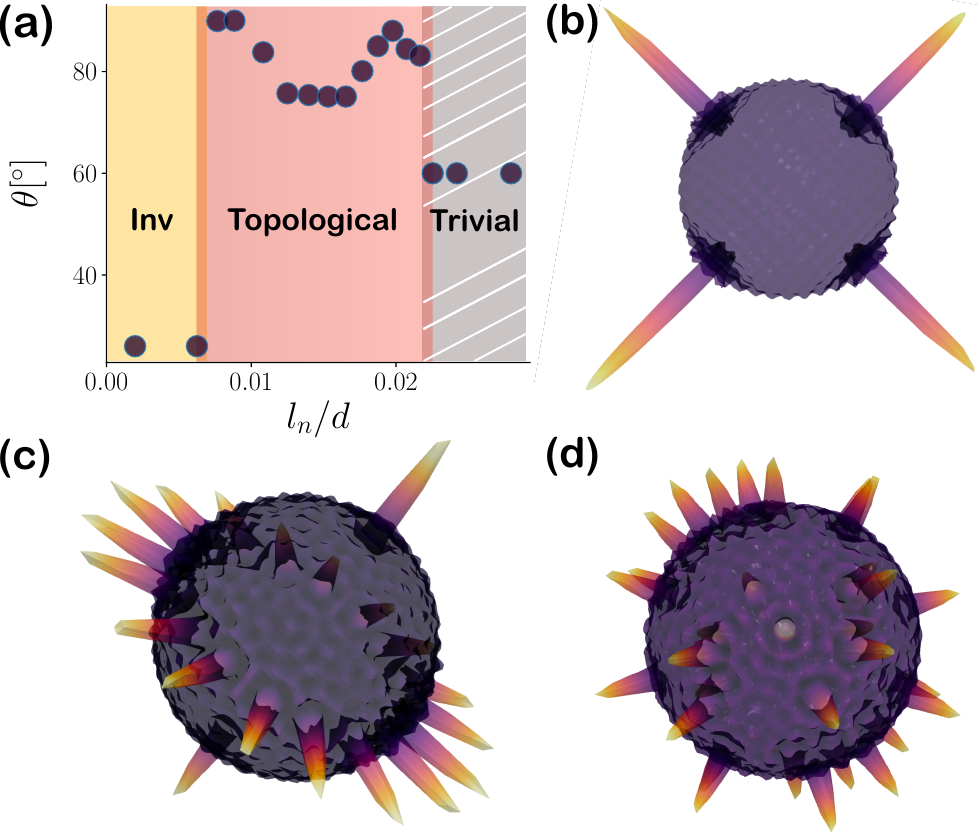}
    \caption{(a) Angle between square bivectors after an exchange, $\theta$, as a function of $l_n/d$.
    Shadings denote the regimes discussed in the text (Inv = Invariant). 
    (b) Superposition of spherical harmonics (up to $99$th order) in a spherical harmonics expansion of the orbit on the Bloch sphere, of a configuration starting from the state in Fig.~\ref{fig1}, following repeated exchange between defects $(1,2)$ (see Fig.~\ref{fig1}), for 
    $l_n\simeq 0.14$. (c,d) Corresponding representation for an orbit of a configuration which is braided according to the pattern (braid word) BDACBDACBDACBDAC (c) and  ABACABACACACACAC (d), where A, B, C, D denote exchanges between particles $(1,2)$, $(1,4)$, $(3,4)$, $(2,3)$ respectively.  
    }     
    \label{fig4}
\end{figure}

In our Majorana square, the role of the $\sigma$ anyons is played by the defects of winding number 
$\pm 1/2$, which can be algebraically represented by $\pm e_{21}$~\cite{head2024}. 
Due to the elastic locking, our Majorana square can also be described by a single qubit, as for Ising anyons. 
More quantitatively, 
Fig.~\ref{fig4}(a) quantifies how the angle between defect square bivectors following successive exchanges varies with the elastic constant. Importantly, only if $l_n/d$ is between a lower and an upper critical threshold ($0.007 < l_n/d< 0.0225$) do we observe a rotation of the defect bivector 
upon exchanging defects $1$ and $2$ (we call this the ``topological'' regime). 
If $l_n/d$ is too small, the elastic interactions are not enough to rotate the defect bivectors upon exchange (Suppl. Movie 2 for $l_n/d=0.006$; we call this the ``invariant'' regime). Conversely, if $l_n/d$ is too large, the defects in the laser traps get too close to each other during the braiding, and annihilate leaving a non-topological defect-free state (Suppl. Movie 3 for $l_n/d=0.023$; we call this the ``trivial'' regime). The topological regime also requires the director field to twist out of the plane. Accordingly, simulations where we enforced a purely 2D director field profile remain in the invariant regime 
(Suppl. Movie 4). 

For ideal Majorana braiding $\theta=90^{\circ}$, whereas for nematic bits in the topological regime $\theta$ depends on $l_n$, approaching $\simeq 90^{\circ}$ for values of $l_n/d$ close to the transition to the invariant regime, such as $l_n/d\simeq 0.009$. For this value of $l_n/d$, we analyse in Fig.~\ref{fig4}(b) the orbit of the defect bivectors on the Bloch sphere following repeated braiding of two defects along one fixed side of the defect square, which alternates between the poles and two antipodal points on the equator with negligible noise, as would occur when braiding Majorana zero modes~\cite{SI}. Instead, braiding following a different regular pattern of exchanges [a ``braid word'', Fig.~\ref{fig4}(c,d)], or a random pattern of exchanges~\cite{SI} leads to more complex orbits, unlike ideal orbits of Majorana braids which can at most visit the $24$ points of the octahedral group~\cite{SI}. We interpret the rotation angle as comprising a geometric contribution, which approaches $90^{\circ}$ in the adiabatic limit, and a dynamical phase~\cite{bernevig2013}, which depends on both elasticity and kinetic effects. Accordingly, the topological regime where $\theta\simeq 90^{\circ}$ extends to lower values of $l_n/d$ if braiding is slowed down~\cite{SI}. 
The presence of a dynamical phase is reminiscent of corrections observed in practical realisations of Majorana braiding~\cite{hodge2025}.

In summary, we have studied the behaviour of a nematic Majorana defect square, where four defects of alternating winding number 
are created in a 2D system by simulating the action of a laser, which locally melts the nematic. 
In our system, elastic interactions lock the directions of their defect bivectors -- the axis around which the defect profile rotates. 
As a result, we can effectively describe the whole defect square by means of a single defect bivector on the Bloch sphere, 
as in a single qubit. 

Our main result is that, when two of the defects in the square are exchanged, they behave as non-Abelian 
anyons in a large range of parameter space. 
Specifically, when two defects of different winding number 
are exchanged,  we 
find that the defect bivector of the system rotates on the Bloch sphere, 
mirroring the nontrivial braiding of non-Abelian anyons, but in a purely classical system. Mechanistically, this results hinges on the fact that rotations of the square bivector on the Bloch hemisphere connects states which are degenerate, or have the same free energy, provided we work in, or near, the one elastic constant approximation~\cite{SI}. The Bloch rotations vary with nematic elasticity, allowing more tunability than possible with ideal Majorana modes, where braiding only generates a discrete point group. 


Our results show that braiding provides a general and robust way to implement quantum-like computations in liquid crystals, 
and we hope this will prove useful for 
topological computation in the future. There are several research directions which our results point to. For instance, it would be interesting to characterise how entangled photons and quantum light physically interact with the Majorana square~\cite{sultanov2024}. Additionally, cholesteric defects and dislocations~\cite{johnson2025,wu2025}, or 3D nematic loops~\cite{head2024,negro2025} may lead to different braiding behaviour and computational gates, which would be interesting to explore.

We thank T. Winyard for useful discussion. AIT received funding from the European Union's Horizon 2020 research and innovation programme under the Marie Marie Sk\l odowska-Curie grant agreement No 101024548.

\newpage

\onecolumngrid

\section{Supplemental Material}
\appendix

\section{Liquid crystal modelling}

In this Section we give more details about the simulations, including the braiding protocol and parameter list. We also present additional results complementing the ones reported in the main text, and we discuss the effect of hydrodynamic flow.

\subsection{Simulation details and parameter list}

To solve the equations of motion for the orientational order in the absence of flow, Eq.~(2) in the main text, we have used a finite difference scheme. The main temperature dependence of the free-energy parameters is in the parameter $\chi$. More specifically, we assume, as is commonly done in the literature, that the coefficient of the quadratic term in the free energy, $\tfrac{A_0}{2}\left(1-\tfrac{\chi}{3}\right)$ can change sign, and depends on temperature as $c(T-T^*)$, where $c>0$ is a constant with appropriate dimensions and $T^*$ is a temperature below which the isotropic phase is unstable and below which it is either metastable or stable (so $T^*$ is a spinodal point). 

Parameters used for simulations were: $\Gamma=0.33775$, $A_0=1$, $\chi=\chi_1=3$ for the liquid crystal outside the laser wells, $\chi=\chi_0=0$ inside the laser wells. The values of $K$ were varied as discussed in the main text, where parameters are given in terms of the nematic correlation length $l_n=\sqrt{K/A_0}$. Simulations shown in the main text were performed within a square lattice with size ${\mathcal L}=32$, with the four wells initially positioned at $(8,24)$, $(8,8)$, $(24,8)$ and $(24,24)$, for defects 1, 2, 3 and 4 respectively. The size of each well was set to $2$ simulation units. 

To simulate adiabatic braiding, we have moved the defects with a small velocity such that they traversed a counterclockwise contour resulting in their exchange. Specifically, to exchange defects $1$ and $2$ (corresponding to the simulations in Fig.~1 in the main text), the trajectories of these two defects were as follows. The centre of the laser well trapping defect $1$ started from position $(x_1,y_1)=(8,24)$, the trap for defect $2$ from $(x_2,y_2)=(8,8)$. The centres of these two wells then evolved for $0\le t \le 500000$ simulation units as follows:
\begin{eqnarray}
    \frac{d}{dt}\begin{pmatrix} x_1(t)\\y_1(t)\end{pmatrix} & = & \begin{pmatrix} -v \\ 0 \end{pmatrix}, \; \; \; 
    \frac{d}{dt}\begin{pmatrix} x_2(t)\\y_2(t)\end{pmatrix}  =  \begin{pmatrix} v \\ 0 \end{pmatrix}\qquad {\rm if \, t_1\le t \le t_2} \\ \nonumber
    \frac{d}{dt}\begin{pmatrix} x_1(t)\\y_1(t)\end{pmatrix} & = & \begin{pmatrix} 0 \\ -v \end{pmatrix}, \; \; \; 
    \frac{d}{dt}\begin{pmatrix} x_2(t)\\y_2(t)\end{pmatrix}  =  \begin{pmatrix} 0 \\ v \end{pmatrix} \; \qquad {\rm if \, t_3\le t \le t_4} \\ \nonumber
    \frac{d}{dt}\begin{pmatrix} x_1(t)\\y_1(t)\end{pmatrix} & = & \begin{pmatrix} v \\ 0 \end{pmatrix}, \; \; \; \; \; \;
    \frac{d}{dt}\begin{pmatrix} x_2(t)\\y_2(t)\end{pmatrix}  =  \begin{pmatrix} -v \\ 0 \end{pmatrix}\; \; \; \; \;{\rm if \, t_5\le t \le t_6} \\ \nonumber
    \frac{d}{dt}\begin{pmatrix} x_1(t)\\y_1(t)\end{pmatrix} & = & \begin{pmatrix} 0 \\ 0 \end{pmatrix}, \; \; \;  \; \; \;
    \frac{d}{dt}\begin{pmatrix} x_2(t)\\y_2(t)\end{pmatrix}  =  \begin{pmatrix} 0 \\ 0 \end{pmatrix}\qquad {\rm otherwise},
\end{eqnarray}
where $v=10^{-4}$ simulation units, $t_1=10000$, $t_2=60000$, $t_3=110000$, $t_4=260000$, $t_5=310000$, $t_6=360000$ simulation units. We verified that performing the braiding by taking twice the time to do each of the exchanges led to similar results.

\subsection{Initial state and ground-state degeneracy for a Majorana square}

Consider ${\mathbf r}_i=(x_i,y_i)$ with $i=1,2,3,4$ the initial positions of the four laser wells, each of which traps a topological defects. These are chosen symmetrically, with the numbering as in Fig.~1 of the main text,
\begin{eqnarray}
    (x_1,y_1)& = &\left(\frac{{\mathcal L}-d}{2},\frac{{\mathcal L}+d}{2}\right), \;
    (x_2,y_2) = \left(\frac{{\mathcal L}-d}{2},\frac{{\mathcal L}-d}{2}\right), \; \\ \nonumber
    (x_3,y_3)& =& \left(\frac{{\mathcal L}+d}{2},\frac{{\mathcal L}-d}{2}\right), \;
    (x_4,y_4)=\left(\frac{{\mathcal L}+d}{2},\frac{{\mathcal L}+d}{2}\right), \; .
\end{eqnarray}
To initialise the liquid crystal pattern, we defined the following angles,
\begin{eqnarray}
    \phi_i (x,y) & = & {\rm atan2}(y-y_i,x-x_i), \\ \nonumber
    \theta_1 & = & -\frac{\phi_1-\pi}{2}, \; \theta_2 = \frac{\phi_2+\pi}{2}
    \\ \nonumber
    \theta_3 & = & -\frac{\phi_3}{2}, \; \theta_4 = \frac{\phi_4}{2}
\end{eqnarray}
and set ${\mathbf n}=(\cos(\theta),\sin(\theta),0)$, where $\theta$ was chosen for each given point of the lattice according to the closest defect to that given point, or, in formulas:
\begin{eqnarray}
    \theta & = & \theta_1 \; {\rm if \; x< \frac{{\mathcal L}}{2}}, \; y\ge  \frac{{\mathcal L}}{2}; \;  \theta =  \theta_2 \; {\rm if \; x< \frac{{\mathcal L}}{2}}, \; y< \frac{{\mathcal L}}{2}; \\ \nonumber
\theta & = & \theta_3 \; {\rm if \; x\ge  \frac{{\mathcal L}}{2}}, \; y< \frac{{\mathcal L}}{2}; \;  \theta =  \theta_4 \; {\rm if \; x\ge \frac{{\mathcal L}}{2}}, \; y \ge \frac{{\mathcal L}}{2}.
\end{eqnarray}
The ${\mathbf Q}$ tensor is then set equal to the uniaxial approximation
\begin{eqnarray}
    Q_{\alpha\beta} & = & q\left(n_{\alpha}n_{\beta}-\frac{\delta_{\alpha\beta}}{3}\right) \\ \nonumber
    q(\chi) & = & \frac{1}{4}+\frac{3}{4}\sqrt{1-\frac{8}{3\chi}},
\end{eqnarray}
where $q(\chi)$ is the magnitude of order expected in thermodynamic equilibrium for a given value of $\chi$.

From this initial condition, the system rapidly evolves to the state shown in Fig.~1 of the main text. This is a sufficiently deep local minimum to correspond to the ground state of our system for all purposes. This ground state has a trivial $S^1\simeq U(1)$ degeneracy, as it is symmetric under global in-plane rotation by any angle (i.e., any angle around the $z$ axis). There is also a subtler degeneracy, associated with the rotation of the defect bivectors on the Bloch hemisphere, which is revealed by braiding the quasiparticles, as the free energy is the same after each exchange: an example is shown by the free energy plot in Fig.~2 of the main text, but this behaviour is more general, and observed following each adiabatic exchange between quasiparticles. This degeneracy is exact in the one elastic constant approximation and is important to allow our liquid crystalline defects to reproduce the behaviour of Ising anyons upon braiding.

\subsection{Spherical harmonics expansion}
\label{sec:si_odf}

To characterise and visualize the orbits on the Bloch sphere  we construct a
smoothed distribution function on the unit sphere and
visualise it as a radially modulated surface.  The procedure is as follows.
We first compute the square Bloch bivector $\mathbf{u}_i$ extracted from the simulation trajectory and duplicate it as its antipode $-\mathbf{u}_i$.  The  points set of the entire orbit is clustered with the DBSCAN algorithm~\cite{dbscan} (neighbourhood radius $\varepsilon = 0.1$, minimum cluster size 3) and each cluster is replaced by its centroid.
The cluster centroids are then recentred (the mean position is subtracted) and
normalised to unit length, yielding a set of $K$ unit vectors
$\{\hat{\mathbf{p}}_k\}$.  Inversion symmetry is enforced a second time on
this final set: for every $\hat{\mathbf{p}}_k$ the antipode
$-\hat{\mathbf{p}}_k$ is added and near-duplicate vectors (coinciding within a
tolerance of $10^{-7}$) are removed.

Rather than binning the points onto a latitude--longitude grid and correcting
for the solid-angle element $\sin\theta$, which introduces a numerical
singularity at the poles, we build the spherical harmonic coefficients directly
from the point positions.  Each unit vector at colatitude $\theta_k$ and
azimuth $\varphi_k$ is treated as a $\delta$-function on the sphere, giving
\begin{equation}
  c_{lm} = \sum_{k=1}^{K} Y_{lm}(\theta_k,\,\varphi_k)\,,
  \label{eq:clm}
\end{equation}
where $Y_{lm}$ are the real, $4\pi$-normalised spherical harmonics.  The
coefficients are evaluated up to degree $l_{\max} = 99$ and subsequently
low-pass filtered by discarding all contributions with $l > l_{\mathrm{cut}} =
40$.   The smoothed density field is then reconstructed on
a Driscoll--Healy grid~\cite{DRISCOLL1994202} as
\begin{equation}
  f(\theta,\varphi)
    = \sum_{l=0}^{l_{\mathrm{cut}}}
      \sum_{m=-l}^{l} c_{lm}\,Y_{lm}(\theta,\varphi)\,.
  \label{eq:odf}
\end{equation}

The smoothed field $f(\theta,\varphi)$ is mapped onto a three-dimensional
surface by modulating the radius of a unit sphere:
\begin{equation}
  R(\theta,\varphi)
    = R_0 + A\,\tilde{f}(\theta,\varphi)\,,
  \label{eq:surface}
\end{equation}
where $\tilde{f} = (f - \langle f \rangle)/\sigma_f$ is the $z$-scored field,
$R_0 = 1$ is the base radius, and $A = 0.15$ controls the deformation
amplitude.  The resulting surface is exported as a VTK structured grid for
rendering in ParaView~\cite{paraview}. 

All spherical harmonic operations were performed with \textsc{pyshtools}~\cite{pyshtools}.

\subsection{Additional results}

In the main text, we showed results corresponding to regular exchanges, or braiding: either repeated braiding of defects $1$ and $2$ (see above and Fig.~1 in the main text for defect labelling), or specific braid words which correspond to a predefined set of exchanges.

It is also of interest to ask what the result of random braiding is. When braiding ideal Majorana anyons (zero modes), as discussed in the analytical Section below, 
the orbit on the Bloch sphere consists of at most $24$ points. In the case of our liquid crystal quasiparticles, elastic effects and small errors due to the angle between successive defect bivectors not being exactly $90^{\circ}$ enlarge the Bloch orbit, as shown in Fig.~S1 (see also Suppl. Movie 6). Nevertheless, the free energy returns to the same value following each exchange, as in the results shown in the main text. The orbit does not appear random though, as it is clear that some points on the sphere are visited more often. Including some random sections in an otherwise regular braiding protocol would increase the number of logic gates which can be achieved, at the price of an increased uncertainty in the state of the system. It is possible that by optimising the duration and location of random events the potential for topological computation is therefore increased rather than hindered; we leave this interesting possibility to future investigation. 

Another relevant generalisation of the results shown in the main text is to defect geometries which are not squares. To address this point, Fig.~S2 (see also Suppl. Movie 7) shows the orbit of the defect bivector characterising a Majorana rectangle where the laser wells are initially positioned at 
\begin{eqnarray}
    (x_1,y_1)& = &\left(\frac{{\mathcal L}_1-d_1}{2},\frac{{\mathcal L}_2+d_2}{2}\right), \; (x_2,y_2)=\left(\frac{{\mathcal L}_1-d_1}{2},\frac{{\mathcal L}_2-d_2}{2}\right), \\ \nonumber
    (x_3,y_3)& = &\left(\frac{{\mathcal L}_1+d_1}{2},\frac{{\mathcal L}_2-d_2}{2}\right), \; (x_4,y_4)=\left(\frac{{\mathcal L}_1+d_1}{2},\frac{{\mathcal L}_2+d_2}{2}\right),
\end{eqnarray}
with ${\mathcal L}_1=2{\mathcal L}_1=64$, and $d_1=2d_2=32$. The orbits we consider in Fig.~S2 correponds to (a) the repeated exchanges of defects $1$ and $4$, and (b) random braiding of two defects along a randomly chosen side of the rectangle.

Finally, in our system there is a finite timescale for defects to reorient their defect bivectors, which can be estimated as the typical timescale for the director field to reorient, $\gamma_1 l^2/K$, where $\gamma_1$ is the rotational viscosity, $K$ is the elastic constant and $l$ is a typical lengthscale, which is likely the system size in our case, as all defects reorient synchronously. If braiding is not sufficiently slow with respect to this elastic timescale, we would expect that reorientation cannot occur. We interpret this as the mechanism leading to the invariant regime in Fig.~4 of the main text. Accordingly, simulations with twice slower braiding show that the topological regime extends to significantly lower values of $l_n/d$, as well as to some subtler changes in the angle between bivectors at successive exchanges at larger $l_n/d$ (Fig.~S3). 

\begin{figure}
\includegraphics[width=12cm]{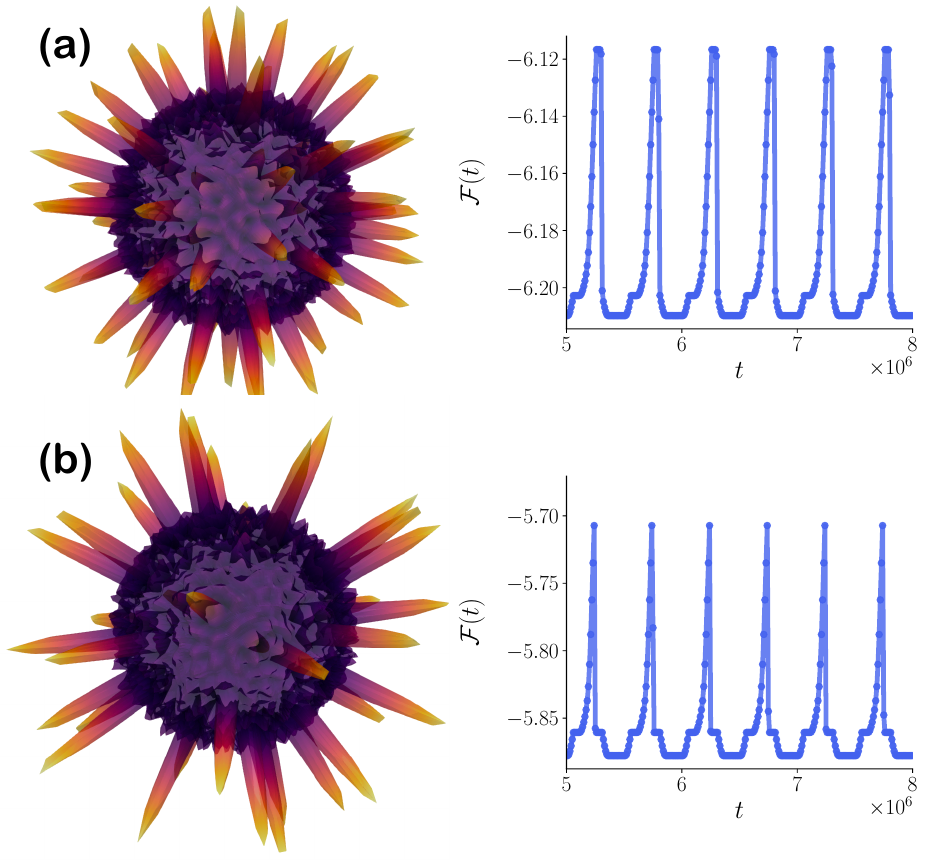}
\caption{(a) Spherical harmonics representation of the Bloch sphere orbit (i) and free energy versus time [(ii); part of the time series only shown for clarity] for random braiding for $K=0.04$. It can be seen that the free energy following each exchange (every $500000$ time steps) is degenerate. (b) Corresponding representation of the Bloch sphere orbit (i) and free energy versus time [(ii); part of the time series only shown for clarity] for $K=0.1$. For both (a) and (b), other parameters are as listed in the text. }
\end{figure}

\begin{figure}
\includegraphics[width=10cm]{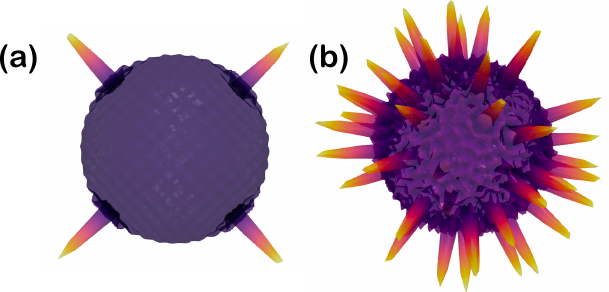}
\caption{Representation of the Bloch sphere orbit for (a) repeated exchanges of defects $1$ and $4$ and (b) random braiding, for a Majorana rectangle (see text for details on the geometry). The elastic constant is $K=0.04$, other parameters as listed in the text. }
\end{figure}

\begin{figure}
\includegraphics[width=8cm]{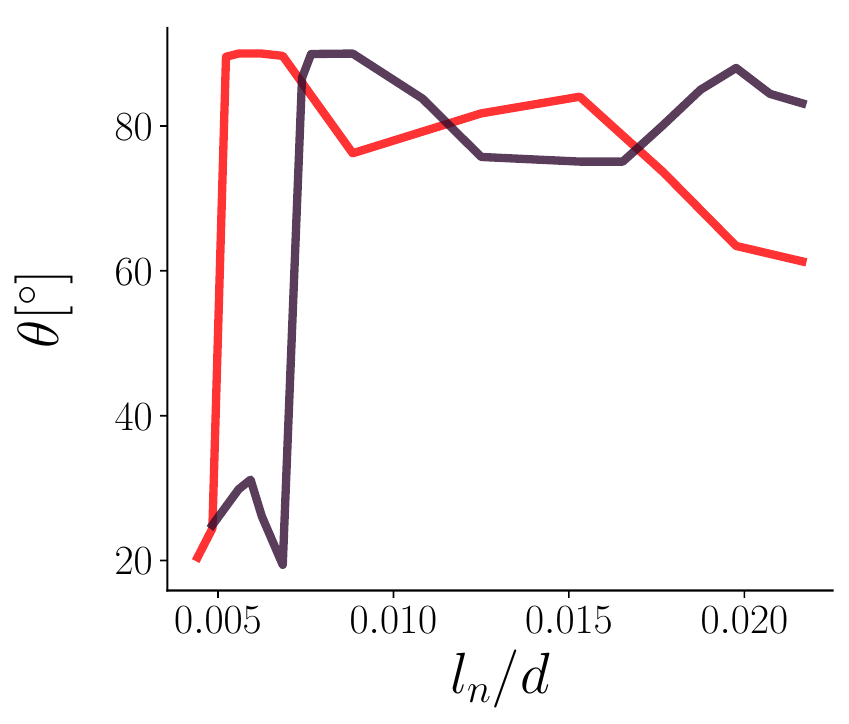}
\caption{Angle between square bivectors after an exchange, $\theta$, as a function of $l_n/d$, for $A_0=1$, and braiding velocity (see text) equal to $v=10^{-4}$ (purple, as in the main text), or $v=5\times 10^{-5}$ (red). The latter, slower, braiding velocity leads to an enlargement of the topological regime.}
\end{figure}

\subsection{Liquid crystal hydrodynamics}

In the main text, we showed results for the case without flow, where the dynamics of the ${\mathbf Q}$ tensor is only due to the molecular field which provides a generalised force  driving the system towards the closest local minimum. In a liquid crystal sample, flow is also generally present and modifies the equations of motion. In this Section we review the equations of motion which describe the dynamics of the ${\mathbf Q}$ tensor describing orientational order, and of the coupled velocity field ${\mathbf v}$. For selected cases, as discussed below, we have solved these more complicated equations, again with hybrid lattice Boltzmann, obtaining qualitatively similar results with respect to those reported in the main text.

The liquid crystal hydrodynamic model we consider is the Beris-Edwards model~\cite{beris1994}. 
The equations of motion governing the dynamics of ${\mathbf Q}$ and ${\mathbf v}$ in this model are the following
\begin{eqnarray}
& & D_t{\bm Q}= {\bm S}({\bm W},{\bm Q}) + \Gamma {\bm H} \; , \label{eqn:7} \\ 
& & \rho (\partial_t + \bm{v}\cdot \nabla){\bf v} = \nabla\cdot ({\bm \sigma}^{\rm hydro}+{\bm \sigma}^{\rm LC}) \; . \label{eqn:8} 
\end{eqnarray}

Eq.~(\ref{eqn:7}) determines the evolution of the orientational tensor order parameter ${\mathbf Q}$. The molecular field ${\mathbf H}=-\left[\frac{\delta {\mathcal F}}{\delta {\bm Q}}+({\bm I}/3){\rm Tr}\frac{\delta {\mathcal F}}{\delta {\bm Q}}\right]$ provides a forcing term which tends to drive the system towards the closest thermodynamic local minimum.  
The term $\bm{S}(\bm{W},\bm{Q})$ in Eq.~(\ref{eqn:8}) denotes the co-rotational derivative, which determines how the flow field rotates and stretches the order parameter, or equivalently determines the contribution of flow beyond advection, which is due to the coupling between the velocity gradient and a tensorial order parameter.  

The explicit expression of the corototational derivative $\bm{S}({\bm W},{\bm Q})$ is
\begin{eqnarray}
\bm{S}({\bm W},{\bm Q}) & = & (\xi{\bm D}+{\bm \Omega})({\bm Q}+{\bm I}/3)\\ \nonumber
& + & (\xi{\bm D}-{\bm\Omega})({\bm Q}+{\bm I}/3) \\ \nonumber
& - & 2\xi({\bm Q}+{\bm I}/3) {\rm Tr} ({\bm Q}{\bm W}). \; .
\label{eqnA1}
\end{eqnarray}
Here, ${\bm D}=({\bm W}+{\bm W}^T)/2$ and ${\bm\Omega}=({\bm W}-{\bm W}^T)/2$ denote the symmetric and anti-symmetric part of the velocity gradient tensor $W_{\alpha\beta}=\partial_{\beta}v_{\alpha}$, respectively. The flow alignment parameter $\xi$ determines the aspect ratio of the LC molecules and the dynamical response of the LC to an imposed shear flow. In this work, we choose $\xi = 0.7$, which corresponds to flow-aligning rod-like molecules. 


Eq.~\eqref{eqn:8} is the Navier-Stokes equation, this determines the evolution of the flow field. Note that we assume incompressibility and a constant (more precisely nearly constant in lattice Boltzmann)  density $\rho$. The stress tensor ${\bm sigma}=({\bm \sigma}^{\rm hydro}+{\bm \sigma}^{\rm LC})$ has been divided in two terms. First, there is a hydrodynamic contribution
\begin{equation}
{\bm \sigma}^{\rm hydro}= - P \bm{I} + \eta \nabla \bm{v},
\end{equation}
which accounts for the hydrodynamic pressure $P$ ensuring incompressibility, and for viscous effects, proportional to the viscosity $\eta$, which is set to $\eta=5/3$ in our simulations. Second, there is a liquid crystal contribution ${\bm \sigma}^{\rm LC}$, which accounts for both elastic and flow-aligning effects.  The explicit expression of the liquid crystalline contribution is 
\begin{eqnarray}
\sigma_{\alpha\beta}^{LC}=&&-\xi H_{\alpha\gamma}(Q_{\gamma\beta}+\frac{1}{3}\delta_{\gamma\beta})-\xi(Q_{\alpha\gamma}+\frac{1}{3}\delta_{\alpha\gamma})H_{\gamma\beta}\\ \nonumber
&&+2\xi(Q_{\alpha\beta}+\frac{1}{3}\delta_{\alpha\beta})Q_{\gamma\mu}H_{\gamma\mu}+Q_{\alpha\gamma}H_{\gamma\beta}-H_{\alpha\gamma}Q_{\gamma\beta}\\ \nonumber
&&-\partial_{\alpha}Q_{\gamma\mu}\frac{\partial f}{\partial(\partial_{\beta}Q_{\gamma\mu})}.
\end{eqnarray}

\begin{figure}
\includegraphics[width=18cm]{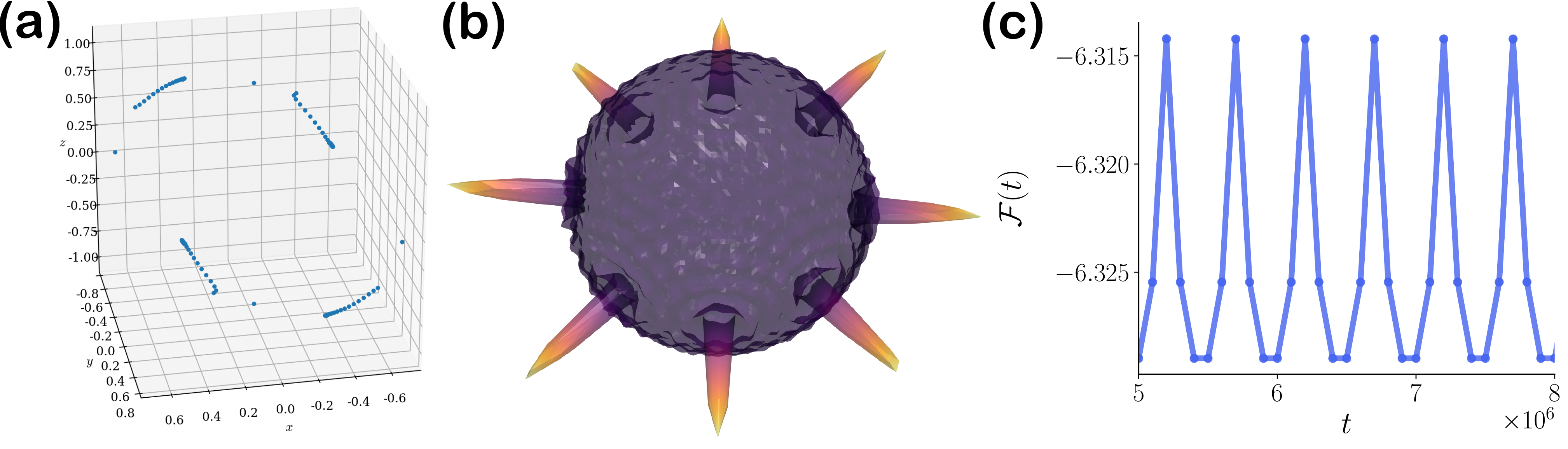}
\caption{(a) Bloch sphere orbit, (b) corresponding spherical harmonics representation and (c) free energy versus time (part of the time series only shown for clarity) corresponding to repeated exchange of defects $1$ and $2$ for $A_0=1$ and $K=0.02$, when flow is included.}
\end{figure}

Fig.~S4 show the Bloch sphere orbit and the free energy versus time for a representative case with flow included, which show that the results discussed in the main text are qualitatively similar with flow, although the average rotation of the defect bivector is further away from $90^{\circ}$ with respect to the case without flow  for most of the parameters we have considered.

\section{Representation of the Braid group $\&$ encoding a qubit with four Majoranas}

\subsection{Fusion rules for Ising anyons}
The fractional quantum Hall state at filling factor $\nu=5/2$, also called the Moore--Read or Pfaffian state, is one of the most promising systems believed to host non-Abelian anyons, in particular Ising anyons \cite{Tong_lecture_notes_16}. Such particles are also the elementary excitations of the Kitaev quantum spin liquid that is thought to be realized, e.g., in $\alpha$-RuCl$_3$ \cite{Do_17}. 
The non-trivial fusion rules for the three types of Ising anyons, namely, the vacuum state ($\1$), a fermion ($\psi$) and an anyon ($\sigma$) are 
\bea
\sigma\times \sigma& =& \1 + \psi\label{eq:sigma_by_sigma}\\
\sigma\times\psi& = &\sigma\\
\psi\times \psi&=& \1\,.
\eea
From the fusion rules, one can deduce the {\it quantum dimension}, $d_\alpha$ with $\alpha\in\left\{\1,\psi,\sigma\right\}$, of the three types of anyons, assuming $d_\1=1$.
\subsection{The Fock space of four Majorana zero modes} 
Majorana zero modes 
are non-Abelian Ising anyons \cite{Hassler_lecture_notes_14}. 
Four ($N$) Majorana zero modes 
at the same energy, created/annihilated by the self-adjoint operators $\,\gamma_i\,=\,\gamma_i^\dagger\,$, generate a Clifford algebra
\bea
\left\{\gamma_i,\gamma_j\right\}&=&2\delta_{ij}\,,
\eea
and span a four-dimensional ($2^{N/2}$-dimensional) Fock space, since $d_\sigma=\sqrt{2}$. According to fusion rule Eq.\ \eqref{eq:sigma_by_sigma}, the two-dimensional vector space of a pair of Majorana zero modes 
can be decomposed into two orthogonal, one-dimensional subspaces: the vacuum state and a 
fermion state. Thus the basis vectors of the four-Majorana Fock space can be chosen e.g.\ as
\bea\label{eq:orig_basis}
|0\rangle\,,\quad\Psi_L^\dagger\,|0\rangle\,,\quad\Psi_R^\dagger\,|0\rangle\,,\quad\Psi_R^\dagger\Psi_L^\dagger\,|0\rangle
\eea
with the fermion creation operators defined as
\bea
\Psi_L^\dagger&\equiv&\frac 1 2 \left(\gamma_1\,+\,i\,\gamma_2\right)\\
\Psi_R^\dagger&\equiv&\frac 1 2 \left(\gamma_3\,+\,i\,\gamma_4\right)
\eea
that obey the canonical anticommutation relations. Conversely,
\bea
\gamma_1&\equiv&\Psi_L^{}\,+\,\Psi_L^\dagger\\
\gamma_2&\equiv&i\,\left(\Psi_L^{}\,-\,\Psi_L^\dagger\right)\\
\gamma_3&\equiv&\Psi_R^{}\,+\,\Psi_R^\dagger\\
\gamma_4&\equiv&i\,\left(\Psi_R^{}\,-\,\Psi_R^\dagger\right)\,.
\eea
The first state, $|0\rangle$, is the vacuum that is annihilated by both $\Psi_{L/R}$.
\begin{figure}
\includegraphics[width=8cm]{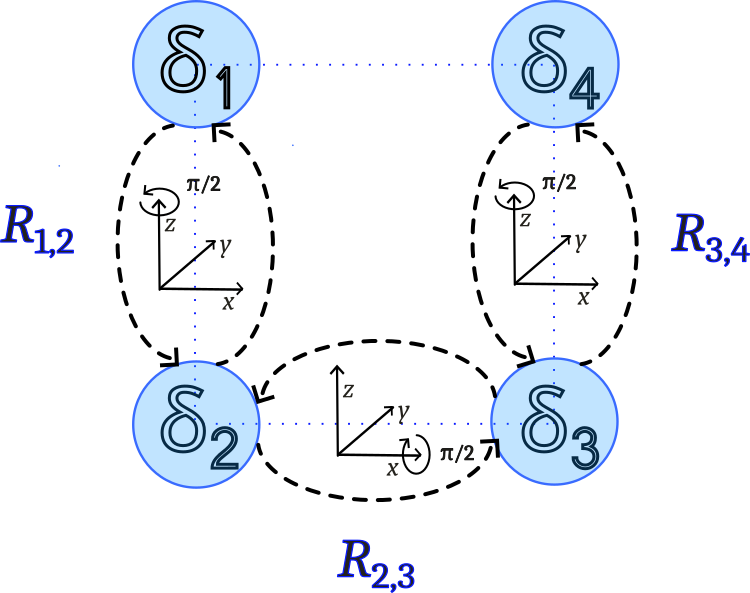}
\caption{Four Majorana zero modes 
located at the vertices of a square. The vector space of a pair of Majorana zero modes 
can be decomposed into two orthogonal subspaces, a vacuum state and a 
fermion state. Exchanging the three pairs of neighboring Majoranas, 
as depicted, generate the braid group $B_4$ which has an infinite number of elements. In the two-dimensional even- and odd-parity subspaces, these nearest neighbour exchanges correspond to $\pi/2$ rotations around the $z$- and $x$-axes when the two-dimensional complex state space is mapped onto a Bloch sphere in three dimensions.\label{fig:Majorana_square}}
\end{figure}

Assuming that the four Majorana zero modes 
are located at the vertices of a square as in Fig.\ \ref{fig:Majorana_square}, exchanging neighboring Majoranas, $i$ and $i+1$ (with $i\in\{1,2,3\}$), in the counter-clockwise direction can be represented by the braiding transformation
\bea
R_{i,i+1}&\equiv&\exp\left(\frac{\pi}{4}\gamma_{i+1}\gamma_{i}\right)\,=\,\frac{1}{\sqrt{2}}\left(\1\,+\,\gamma_{i+1}\gamma_{i}\right)
\eea
that is unitary. The action $\,R_{i,i+1}^{}\,\gamma_j\,R_{i,i+1}^\dagger\,$ results in $\gamma_i\rightarrow\gamma_{i+1}$, $\gamma_{i+1}\rightarrow-\gamma_{i}$, and the other Majoranas remain unaffected. 
The braid group $B_4$ is generated e.g.\ by the following three nearest-neighbor transpositions, $R_{1,2},R_{2,3},R_{3,4}$, that can be rewritten in terms of
the fermion operators as
\bea\label{eq:R12_def}
R_{\,1,2}&=&\frac{1}{\sqrt{2}}\left[
  \left(1\,+\,i\right)\1\,-\,i\,2\,\Psi_L^\dagger\Psi_L^{}\right]\,=\,e^{i\frac{\pi}{4}}\,\1\,-\,i\,\sqrt{2}\,\Psi_L^\dagger\Psi_L^{}\\\label{eq:R23_def}
R_{\,2,3}&=&\frac{1}{\sqrt{2}}\left[\1\,+\,i\,\left(\Psi_R^{}\,+\,\Psi_R^\dagger\right)\,\left(\Psi_L^{}\,-\,\Psi_L^\dagger\right)\right]\,=\,\frac{1}{\sqrt{2}}\left[\1\,+\,i\,\left(\Psi_L^\dagger\Psi_R^{}\,+\,\Psi_L^\dagger\Psi_R^{\dagger}\,+\,\Psi_R^\dagger\,\Psi_L^{}\,+\,\Psi_R^{}\,\Psi_L^{}\,\right)\right]\\\label{eq:R34_def}
R_{\,3,4}&=&\frac{1}{\sqrt{2}}\left[
  \left(1\,+\,i\right)\1\,-\,i\,2\,\Psi_R^\dagger\Psi_R^{}\right]\,=\,e^{i\frac{\pi}{4}}\,\1\,-\,i\,\sqrt{2}\,\Psi_R^\dagger\Psi_R^{}\,.
\eea
On the above four-dimensional Fock space, these operators take the unitary forms
\beq\label{eq:braiding_mxs_nat_bas}
R_{\,1,2}\,=\,
\begin{pmatrix}
e^{i\pi/4} & 0 & 0 & 0 \\
0& e^{-i\pi/4} &0& 0\\
0&0& e^{i\pi/4} & 0\\
0&0&0 & e^{-i\pi/4}
\end{pmatrix}\,,\quad
R_{\,2,3}\,=\,\frac{1}{\sqrt{2}}\,
\begin{pmatrix}
1 & 0 & 0 & -i \\
0&1 &i& 0\\
0&i&1 & 0\\
-i&0&0 &1
\end{pmatrix}\,,\quad
R_{\,3,4}\,=\,
\begin{pmatrix}
e^{i\pi/4} & 0 & 0 & 0 \\
0& e^{i\pi/4} &0& 0\\
0&0& e^{-i\pi/4} & 0\\
0&0&0 & e^{-i\pi/4}
\end{pmatrix}\,,
\eeq
which lend themselves to the following (unitary) representation of the $\gamma_i$s (that are self-adjoint)
\bea
\gamma_1\,=\,
\begin{pmatrix}
0 & 1 & 0 & 0 \\
1&0 &0& 0\\
0&0&0 & -1\\
0&0&-1 &0
\end{pmatrix}
,\quad
\gamma_2\,=\,
\begin{pmatrix}
0 & i & 0 & 0 \\
-i&0 &0& 0\\
0&0&0 & -i\\
0&0&i &0
\end{pmatrix}
,\quad
\gamma_3\,=\,
\begin{pmatrix}
0 & 0 & 1 & 0 \\
0&0 &0& 1\\
1&0&0 & 0\\
0&1&0 &0
\end{pmatrix}
,\quad
\gamma_4\,=\,
\begin{pmatrix}
0 & 0 &i & 0 \\
0&0 &0&i\\
-i&0&0 & 0\\
0&-i&0 &0
\end{pmatrix}\,,
\eea
i.e.\
\bea
\gamma_1\,=\,\sigma_z\otimes\sigma_x\,,\quad\gamma_2\,=\,-\,\sigma_z\otimes\sigma_y\,,\quad\gamma_3\,=\,\sigma_x\otimes\1\,,\quad\gamma_4\,=\,-\,\sigma_y\otimes\1\,.
\eea
that generate the Clifford algebra Cl(4,0). In this representation, 
\bea
\Psi_L^\dagger\,=\,
\begin{pmatrix}
0 & 0 &0 & 0 \\
1&0 &0&0\\
0&0&0 & 0\\
0&0&-1 &0
\end{pmatrix}
,\quad
\Psi_R^\dagger\,=\,
\begin{pmatrix}
0 & 0 & 0 & 0 \\
0&0 &0&0\\
1&0&0 & 0\\
0&1&0 &0
\end{pmatrix}\,.
\eea

\subsection{Subspaces of different fermion parity}
If we assume that the Hamiltonian commutes with the left/right fermion parity operators defined for the left/right pair of Majoranas as
\bea
{\mathcal P}_L\,\,\equiv\,\,i\gamma_1\gamma_2\,,\,\quad\quad\quad{\cal P}_R\,\,\equiv\,\, i\gamma_3\gamma_4\,\\ \nonumber
\eea
and consequently also with the total parity operator
\bea
{\cal P}_{tot}&=&{\mathcal P}_L\,{\mathcal P}_R\,,
\eea
then the Fock space separates into even- and odd-parity subspaces with ${\cal P}_{tot}$ eigenvalues $\pm 1$, respectively.  Namely, the even-parity
subspace spanned by  $\{|0\rangle\,,\,\Psi_R^\dagger\Psi_L^\dagger\,|0\rangle\}$ does not mix with the odd-parity subspace $\{\Psi_L^\dagger\,|0\rangle\,,\,\Psi_R^\dagger\,|0\rangle\}$. Thus,  to encode a qubit, four Majorana zero modes 
are needed. The even-parity subspace of four Majoranas can support a single logical qubit where 0 corresponds to the vacuum state, 1 to the two-fermion state, and any superposition of these two states are also allowed. However, by braiding four Majoranas, we cannot access arbitrary superpositions of these two orthogonal qubit states; only 24 discrete states can be generated, rather than a continuum, as shown below.

\begin{figure}
\includegraphics[width=7cm]{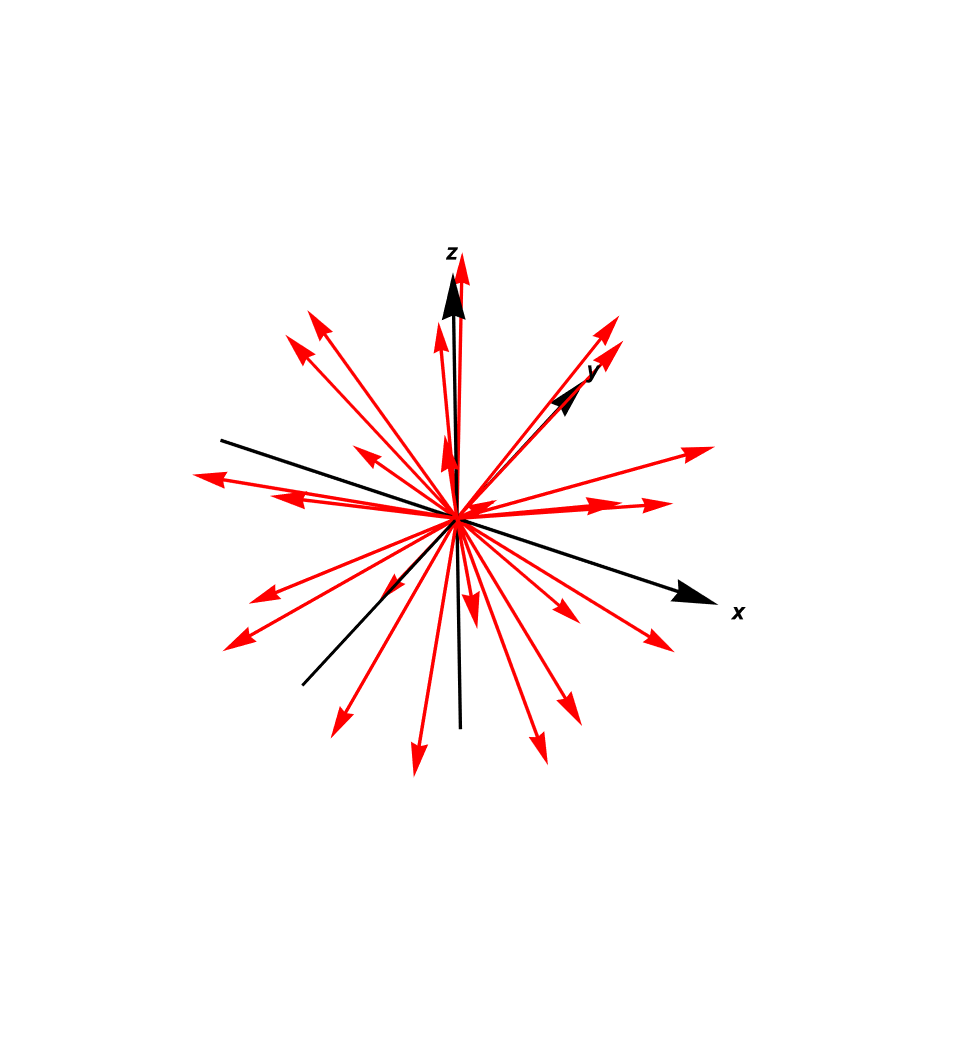}
\includegraphics[width=6cm]{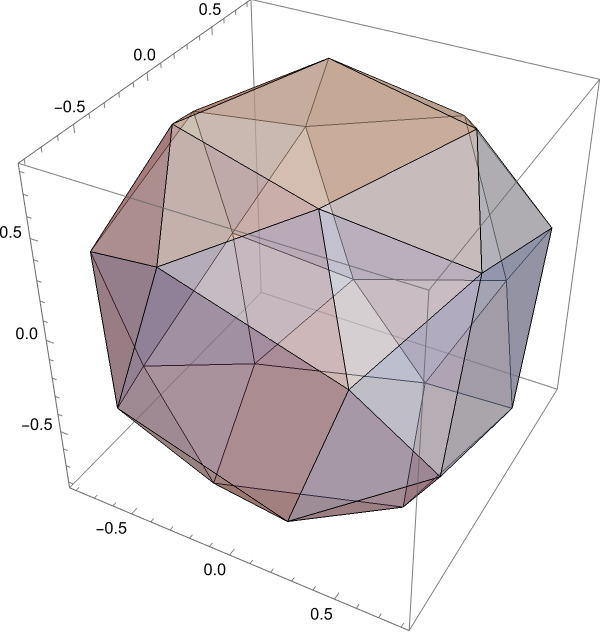}\caption{Orbit of a general, not high-symmetry point under ${\cal O}$, the octahedral group.}
\label{fig:O_group}
\end{figure}

On the even-parity subspace, denoted by the superscript $+$, the braiding operators can be represented as (c.f.\ Eq.\ \eqref{eq:braiding_mxs_nat_bas})
\beq
R_{\,1,2}^{\,+}\,=\,
R_{\,3,4}^{\,+}\,=\,
\begin{pmatrix}
e^{i\pi/4} & 0  \\
0& e^{-i\pi/4}
\end{pmatrix}\,=\exp\left(i\,\frac{\pi}{4}\,\sigma_z\right)\,,\quad
R_{\,2,3}^{\,+}\,=\,\frac{1}{\sqrt{2}}\,
\begin{pmatrix}
1  & -i \\
-i &1
\end{pmatrix}\,=\exp\left(-i\,\frac{\pi}{4}\,\sigma_x\right)\,\,,
\eeq
that correspond to rotations by $\pm\pi/2$ on the Bloch sphere around the $z$- and $x$-axes as depicted in Fig.\ \ref{fig:Majorana_square}.
These two rotations generate the whole octahedral group ${\cal O}$ with 24 elements. To see this, we note that  ${\cal O}$ can be generated by two elements, e.g.\ by the $\pi/2$ rotation around the $z$-axis, and the $2\pi/3$ rotation about the $(1,1,1)$ direction. The latter corresponds to $R_{\,1,2}^{\,+}\left(R_{\,2,3}^{\,+}\right)^3$.  Under ${\cal O}$, the orbit of a generic point on the Bloch sphere that does not lie along any symmetry axis defines a polyhedron with 24 vertices depicted in Fig.\ \ref{fig:O_group}.

\subsection{Movies description}
\begin{itemize}

\item{{\bf Suppl. Movie 1:} Movie corresponding to the full dynamics of the simulation shown in Fig.~2 of the main text, corresponding to the repeated exchange of defects $1$ and $2$, with $A_0=1$, $K=0.04$.}

\item{{\bf Suppl. Movie 2:} Movie corresponding to the dynamics of a simulation corresponding to the repeated exchange of defects $1$ and $2$, with $A_0=1$, $K=0.001$.}

\item{{\bf Suppl. Movie 3:} Movie corresponding to the dynamics of a simulation corresponding to the repeated exchange of defects $1$ and $2$, with $A_0=1$, $K=0.15$, showing the annihilation of the two braided defects.}

\item{{\bf Suppl. Movie 4:} Movie corresponding to the dynamics of a simulation corresponding to the repeated exchange of defects $1$ and $2$, with $A_0=1$, $K=0.04$, with the director field forced to remain in plane. In this case the defect bivector cannot rotate away from $\pm \hat{z}$, hence there is no non-trivial dynamics on the Bloch hemisphere.}

\item{{\bf Suppl. Movie 5:} Movie corresponding to the full dynamics of the simulation shown in Fig.~3 of the main text, corresponding to the repeated exchange of defects $2$ and $4$, with $A_0=1$, $K=0.04$.}

\item{{\bf Suppl. Movie 6:} Movie corresponding to  random braiding with $A_0=1$ and $K=0.04$. With respect to the simulation shown in Fig.~S1, also diagonal exchanges are included here. }

\item{{\bf Suppl. Movie 7:} Movie corresponding to the full simulations shown in Fig.~S2(a), corresponding to repeated exchange of defects $1$ and $4$ in a Majorana rectangle (see text), with $A_0=1$ and $K=0.04$.}



\end{itemize}

\end{document}